\documentclass[9pt,conference]{IEEEtran}
\IEEEoverridecommandlockouts

\usepackage{cite}
\usepackage{amsmath,amssymb,amsfonts}
\usepackage{algorithmic}
\usepackage{graphicx}
\usepackage{textcomp}
\usepackage{xcolor}
\usepackage{amsmath,graphicx}
\usepackage{multirow}
\usepackage{booktabs}
\usepackage{amssymb}
\usepackage{threeparttable}

\def\BibTeX{{\rm B\kern-.05em{\sc i\kern-.025em b}\kern-.08em
    T\kern-.1667em\lower.7ex\hbox{E}\kern-.125emX}}
\begin{document}

\title{Advanced Zero-Shot Text-to-Speech for Background Removal and Preservation with Controllable Masked Speech Prediction
}

\author{\IEEEauthorblockN{Leying Zhang \qquad Wangyou Zhang \qquad Zhengyang Chen \qquad Yanmin Qian$^\dagger$\thanks{$^\dagger$ Corresponding Author}}
\IEEEauthorblockA{\textit{Auditory Cognition and Computational Acoustics Lab, MoE Key Lab of Artificial Intelligence, AI Institute} \\
\textit{Department of Computer Science and Engineering, Shanghai Jiao Tong University, Shanghai, China 
}\\
\{zhangleying, wyz-97, zhengyang.chen, yanminqian\}@sjtu.edu.cn}
}


\maketitle

\begin{abstract}
The acoustic background plays a crucial role in natural conversation. It provides context and helps listeners understand the environment, but a strong background makes it difficult for listeners to understand spoken words. The appropriate handling of these backgrounds is situation-dependent: Although it may be necessary to remove background to ensure speech clarity, preserving the background is sometimes crucial to maintaining the contextual integrity of the speech. Despite recent advancements in zero-shot Text-to-Speech technologies, current systems often struggle with speech prompts containing backgrounds. To address these challenges, we propose a Controllable Masked Speech Prediction strategy coupled with a dual-speaker encoder, utilizing a task-related control signal to guide the prediction of dual background removal and preservation targets. Experimental results demonstrate that our approach enables precise control over the removal or preservation of background across various acoustic conditions and exhibits strong generalization capabilities in unseen scenarios.  
\end{abstract}


\section{Introduction}
\label{sec:intro}

Zero-shot Text-to-Speech (TTS) technology aims to generate natural speech for unseen speakers with speech prompts~\cite{ju2024naturalspeech}. Although recent advancements in deep learning have improved the quality of synthesized speech~\cite {wang2023neural, shen2023naturalspeech,zhang2024covomix}, they still struggle with speech prompts of low quality or polluted by environmental noises, reverberation or interfering speaker~\cite{wang2024investigation}. The difficulty in handling these prompts is two-fold. First, TTS systems should remove unwanted background from the prompt to produce clean speech. Second, when the background in the prompt is important, TTS systems should generate speech with a consistent background. We define this dual requirement as the task of background removal and preservation. The former aims to generate clean speech with the target speaker’s timbre, while the latter involves generating speech that maintains both the speaker's timbre and acoustic background in the prompt.

Most research has focused on background removal, also known as the noise-robust TTS task.  The most straightforward approach involves applying speech enhancement (SE) to the noisy speech prompt~\cite{valentini2018speech, zhang2023toward, zhang2023diffusion}. However, even the state-of-the-art (SOTA) SE models inevitably introduce processing artifacts, and degrade the speech quality produced by the TTS system~\cite{iwamoto2022bad}. Other studies have explored environment conditioning, self-supervised learning, data filtering, and masked speech denoising to achieve clean speech generation in TTS systems~\cite{saeki22_interspeech,zhang2021denoispeech,fujita2024noise,zhao2024experimental}. Although current systems are effective for additional noise, they may still degrade the timbre similarity compared with the speech prompt, especially for prompts with reverberation and interfering speakers.

In contrast, less attention has been paid to background preservation. VoiceLDM~\cite{lee2024voiceldm} and Audiobox~\cite{vyas2023audioboxunifiedaudiogeneration} synthesize speech with acoustic background controlled by text descriptions. 
EATTS~\cite{tan22_interspeech} achieves background preservation only in the reverberation scenario because it is a relatively more stable acoustic characteristic. Related research on the noisy-to-noisy voice conversion~\cite{yao2023preserving, xie2022direct,xie2023noisy}  preserve the exact background of the source speech, but different from the TTS task, these input and output speeches share the same length. Recently, some zero-shot TTS systems,  such as VALL-E~\cite{wang2023neural}, SpeechX~\cite{wang2024speechx} and VoiceBox~\cite{VoiceBox}, have demonstrated potential in generating speech that reflects the acoustic environment of the input speech prompt. 

However, these systems lack a mechanism to controll background preservation, leading to unstable results with different prompts. Moreover, the timbre and background in the prompt are intertwined in systems like VoiceBox,  leading to inconsistencies where the background is only present during spoken segments and absent in silent parts. Furthermore, none of the existing systems can simultaneously remove or preserve environmental information due to the fact that background removal and preservation are inherently conflicting tasks, which can potentially interfere with each other’s effectiveness. 

To address these challenges, we propose a zero-shot TTS system that utilizes a task-related control signal to guide the prediction of dual background removal and preservation targets. The key contributions are as follows, and we invite you to listen to our audio samples\footnote{https://vivian556123.github.io/TTS-for-BR-and-BP}.

1) We investigate the deficiency of current zero-shot TTS systems in the background removal and preservation task, and we propose the Controllable Masked Speech Prediction strategy to enhance the controllability in background removal and preservation. 

2) We introduce the Dual Speaker Encoder that effectively models the speaker timbre and the acoustic background, minimizing interference during multi-task training.

3)  Experimental results demonstrate the effectiveness and controllability of our method against both in-domain and out-of-domain degraded prompts, including those impacted by environmental noise, reverberation, and interfering speakers.

\section{Related Work}

\subsection{Flow-Matching Based Zero-Shot TTS}
\label{sec:related_work}

Continuous Normalizing Flows (CNF) \cite{chen2018neural} is a family of generative models. It parameterizes the time-dependent vector field to construct a flow that learns the transformation from a simple prior distribution (i.e. normal distribution) to the data distribution~\cite{song2021maximum}. 
Flow Matching (FM)~\cite{lipman2023flow} is a simulation-free approach for training CNFs. The relationship between the vector field and the flow is defined via the ordinary differential
equation (ODE). Recent TTS models such as VoiceBox~\cite{VoiceBox}, Audiobox~\cite{vyas2023audioboxunifiedaudiogeneration} and CoVoMix~\cite{zhang2024covomix}  have demonstrated the capability of flow-matching models to generate high-quality speeches. However, these models are sensible for speech prompts, and synthesize speech with degraded quality, especially with noisy speech prompts. 

Our TTS system follows VoiceBox, consisting of an acoustic model and a duration predictor. The acoustic model models the conditional distribution $q(x | z, x_{ctx})$. Specifically, it generates mel-spectrogram $x$ given phoneme sequences $z$ and prompt $x_{ctx}$.  This model serves as a vector field estimator, where at each timestamp $t \in [0,1]$, a randomly chosen masked part $\tilde{x} = x\odot mask$ is predicted, while the visible part $x_{ctx} = x \odot (1-mask)$ is treated as prompt. 
The training objective is described in Eq.\ref{eq:acous}~\cite{VoiceBox},  with the transformer output $v_t(w, x_{ctx}, z; \theta)$, the flow $w$ at step $t$, and the Gaussian noise $x_0$ sampled from the normal distribution. $\sigma_{min}$ is a hyper-parameter to control the deviation of flow-matching.  
\begin{equation}
    \mathcal{L}_{VoiceBox} = \mathbb{E} \Vert (\tilde{x} - (1 - \sigma_{min})x_0) - v_t(w, x_{ctx}, z ; \theta) \Vert^2
    \label{eq:acous}
\end{equation}

VoiceBox~\cite{VoiceBox} employs classifier-free guidance (CFG) to trade off mode coverage and sample fidelity \cite{VoiceBox,ho2022classifier}. 
During training, the acoustic prompt $x_{ctx}$ and phoneme sequences $z$ are dropped with $p_{uncond}$. During inference, VoiceBox first samples a Gaussian noise $x_0$ from normal distribution $p_0$ and uses an ODE solver to evaluate flow $w$. The modified vector field of the CFG strategy becomes $\tilde{v}_t(w, x_{ctx}, z; \theta)$ in Eq.~\ref{eq:cfg} instead of $v_t(w, x_{ctx}, z; \theta)$, where $\alpha$ is a hyperparameter controlling the strength of guidance. 
\begin{equation}
    \tilde{v}_t(w, x_{ctx}, z ; \theta) = (1 + \alpha) v_t(w, x_{ctx}, z ; \theta) - \alpha \tilde{v}_t(w; \theta)
    \label{eq:cfg}
\end{equation}

The duration predictor, similar to the acoustic model, models the conditional distribution $q(y | l, y_{ctx})$, with $y$ the predicted duration sequence of each phoneme, given the input phoneme index sequences $l$ and prompt $y_{ctx}$. 

\subsection{Noise Robust TTS}
Previous research has focused on enhancing the robustness of TTS models in generating clean speech from noisy prompt~\cite{saeki22_interspeech,zhang2021denoispeech,fujita2024noise}. Recently, ~\cite{wang2024investigation} enhances the flow-matching model's noise robustness through masked speech denoising (MSD) strategy, which trains the acoustic model with the context $x_{ctx}$ augmented by noise and predicts the original clean target. However, while effective in scenarios involving additional noise, this method lost the background preservation capabilities inherent in the conventional flow-matching model, and restricts its application in scenarios where maintaining the acoustic environment of the prompt is critical.

\begin{table*}[htbp]
    \centering
    \caption{Evaluation results for background removal. \textbf{Bold} indicates the best result. OOD means the Out-of-domain dataset. }
    \label{tab:removal}
    \setlength\tabcolsep{4.7pt}
    \begin{tabular}{c|ccc|ccc|ccc|ccc|ccc}
    \toprule
        \multirow{2}{*}{System} &  \multicolumn{3}{c|}{Clean} & \multicolumn{3}{c|}{Noise} & \multicolumn{3}{c|}{Reverb} & \multicolumn{3}{c}{Interference} & \multicolumn{3}{c}{VCTK-TUT (OOD)}  \\ 
     ~ & SIM$\uparrow$ & MCD$\downarrow$ & MOS$\uparrow$  &   SIM$\uparrow$  & MCD$\downarrow$ & MOS$\uparrow$ &   SIM$\uparrow$  & MCD$\downarrow$ & MOS$\uparrow$ &   SIM$\uparrow$  & MCD$\downarrow$ & MOS$\uparrow$ &   SIM$\uparrow$  & MCD$\downarrow$ & MOS$\uparrow$ \\ \midrule
Clean GT & 0.68 & / & 4.57 & 0.68 & / & 4.57 & 0.68 & / & 4.57 & 0.68 & / & 4.57 & 0.54 & / & 4.82 \\ \midrule      
        VoiceBox & \textbf{0.59} & 6.28 & \textbf{4.08} & 0.32 & 14.69 & 2.00 & 0.51 & 9.18 & 3.03 & 0.45 & 9.27 & 2.97  & 0.21 & 13.03 & 2.35 \\ 
        VoiceBox+SE & 0.58 & 7.21 & 3.90 & \textbf{0.45} & 7.24 & 3.66 & 0.46 & 8.09 & 3.15 & 0.44 & 7.91 & 3.32 & \textbf{0.33} & \textbf{6.20} & \textbf{3.95}\\ 
        VoiceBox+MSD & 0.54 & 6.33 & 3.94 & 0.39 & 6.86 & 3.78 & 0.47 & 6.72 & 3.88 & 0.46 & 6.62 & 3.98 & 0.27 & 6.70 & 3.90 \\ \midrule
        Ours & \textbf{0.59} & \textbf{6.22} & 3.96 & 0.44 & \textbf{6.66} & \textbf{4.05}& \textbf{0.51} & \textbf{6.65} & \textbf{4.02} & \textbf{0.50} & \textbf{6.56} & \textbf{4.09} & 0.29  & 6.32  & 3.94\\  
        \bottomrule
    \end{tabular}
    \vspace{-0.2cm}
\end{table*}

\begin{table*}[htbp]
    \centering
    \caption{Evaluation results for background preservation. \textbf{Bold} indicates the best result.  OOD means the Out-of-domain dataset. }
    \label{tab:preserve}
    \setlength\tabcolsep{5.2pt}
    \begin{tabular}{c|cc|ccc|ccc|ccc|ccc}
    \toprule
        \multirow{2}{*}{System}&  \multicolumn{2}{c|}{Clean} & \multicolumn{3}{c|}{Noise} & \multicolumn{3}{c|}{Reverb} & \multicolumn{3}{c}{Interference} & \multicolumn{3}{c}{VCTK-TUT (OOD)}  \\ 
     ~ & SIM$\uparrow$ & MCD$\downarrow$ &   SIM$\uparrow$  & MCD$\downarrow$ & BMOS$\uparrow$ &   SIM$\uparrow$  & MCD$\downarrow$ & BMOS$\uparrow$ &   SIM$\uparrow$  & MCD$\downarrow$ & BMOS$\uparrow$ & SIM$\uparrow$  & MCD$\downarrow$ & BMOS$\uparrow$  \\ \midrule
Noisy GT & 0.68 & / & 0.57 & / & 4.72 & 0.63 & / & 4.66 & 0.50 & / & 4.69 & 0.49 & / & 4.74\\ \midrule
         VoiceBox & \textbf{0.59}  & 6.28   & 0.32  & 13.15  & 2.09 & \textbf{0.51}  & 9.30  & 2.60 &  0.45 & \textbf{12.21} & 1.88 & 0.21 & 12.47 & 2.28 \\ 
        Ours & \textbf{0.59}   & \textbf{6.22}   & \textbf{0.43}  & \textbf{11.00}  & \textbf{3.78} & \textbf{0.51}  & \textbf{9.07}  & \textbf{3.51} & \textbf{0.48} & 12.33 & \textbf{3.09} & \textbf{0.29} & \textbf{9.52}  & \textbf{3.98} \\  \bottomrule
    \end{tabular}
        \vspace{-0.2cm}
\end{table*}

\section{Methodology}
\label{sec:method}

To achieve controllable background removal and preservation by giving a speech prompt with environmental information, we propose the Controllable Masked Speech Prediction training strategy and we design a dual speaker encoder to process the speech prompt.

\subsection{Controllable Masked Speech Prediction}

While the MSD approach in \cite{wang2024investigation} effectively estimates clean audio from noisy prompts, it cannot selectively preserve the acoustic environment. To address this, we propose a novel Controllable Masked Speech Prediction (CMSP) strategy for the acoustic model, which simultaneously manages background removal and preservation by uniting these tasks into the masked prediction problem. It allows the model to perform dual-masked predictions within the same training framework. This dual-task capability ensures that the model can flexibly adapt to different application scenarios while maintaining consistent speaker characteristics.

Specifically, with probabilities $P^{N}$, $P^{R}$, $P^{IS}$ for different noise types, we simulate noisy speech $x^{aug}$ by augmenting clean training samples $x$ with environmental noise, reverberation and interfering speech respectively.  The probability of without background adding, i.e. $x^{aug}=x$ is denoted as $P^{C}$. A random portion of the speech is masked and to be predicted, noted as $\tilde{x}$ and $\tilde{x}^{aug}$ for samples before and after augmentation. The remaining unmasked segments are the prompt  $x_{ctx}^{aug}$. We design a multi-task learning loss as shown in Eq.\ref{eq:cmsp} based on the conventional flow-matching loss in Eq.\ref{eq:acous}, with notations described in Section.\ref{sec:related_work}.  By introducing binary control signals $c (c_0\text{ and }c_1)$, the model is guided to focus on either removing or retaining background elements, thus allowing for precise control over the generated speech's acoustic environment.  
\begin{equation}
\label{eq:cmsp}
\begin{aligned}
     \mathcal{L}_{CMSP} & =  \underbrace{\mathbb{E} \Vert (\tilde{x} - (1 - \sigma_{min})x_0) - v_t(w, x_{ctx}^{aug}, z, c_0 ; \theta) \Vert^2}_{\text{Backgroud Removal Loss}} \\
      &+ \underbrace{\mathbb{E} \Vert (\tilde{x}^{aug} - (1 - \sigma_{min})x_0) - v_t(w, x_{ctx}^{aug}, z, c_1 ; \theta)\Vert^2 }_{\text{Backgroud Preservation Loss}}
\end{aligned}
\end{equation}

\vspace{-0.3cm}
\subsection{Dual Speaker Encoder}

\begin{figure}
    \centering
\includegraphics[width=1.0\linewidth]{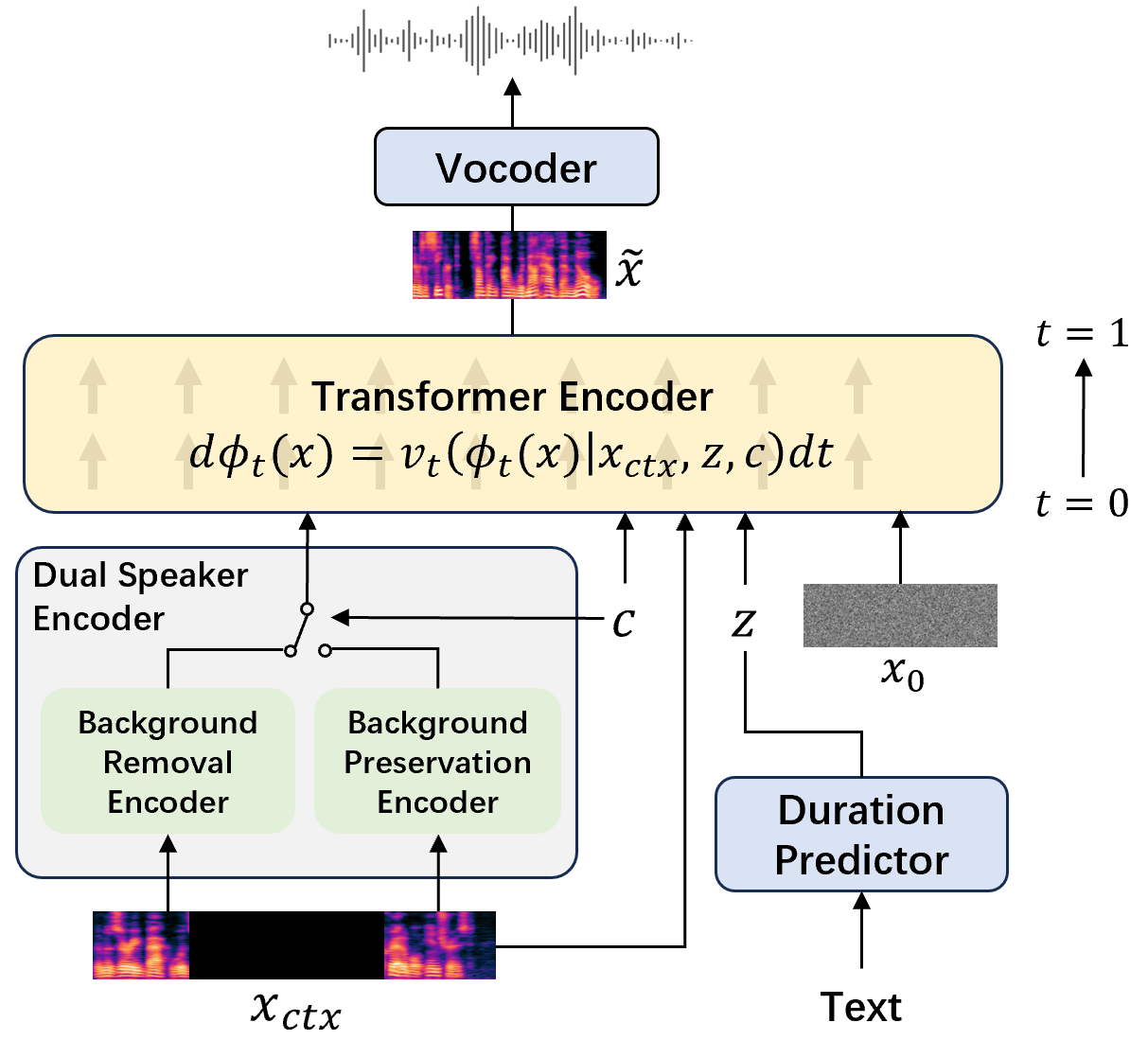}
    \caption{Model Architecture}
    \label{fig:model}
    \vspace{-0.4cm}
\end{figure}
Previous research demonstrates the importance of speaker information robustness when dealing with degraded prompts~\cite{fujita2024noise}. However,  VoiceBox processes noisy prompts by simply concatenating them into the model input, which may limit its ability to accurately extract speaker information under challenging acoustic conditions. To address this limitation, we propose the dual speaker encoder that is specifically designed to better handle noisy prompts. As illustrated in Figure~\ref{fig:model}, in addition to concatenating the noisy prompt, we incorporate two identical encoders into the model, which are fed with the noisy prompt independently. A control mechanism, guided by a binary control signal $c$, determines which branch of the speaker encoder to activate. This design can minimize computational cost in the model interference between the removal and preservation tasks, and it further improves the background preservation performance.

\vspace{-0.1cm}
\subsection{Inference Process}
Given the control signal $c$ which specifies the task to be performed, the inference process begins with the duration predictor generating the phoneme sequence $z$ based on the given transcription. Then, a noisy prompt $x_{ctx}$, the phoneme sequence $z$, a Gaussian noise $x_0$, and the control signal $c$ will be sent to the acoustic model. Similar to the CFG strategy in VoiceBox (Eq.\ref{eq:cfg}), the acoustic model predicts the final mel-spectrogram with an ODE solver guided by the vector field defined in Eq. \ref{eq:cfg_ours}. A HiFiGAN vocoder finally converts the mel-spectrogram into the waveform. 
\begin{equation}
    \tilde{v}_t(w, x_{ctx}, z,c ; \theta) = (1 + \alpha) v_t(w, x_{ctx}, z, c ; \theta) - \alpha \tilde{v}_t(w, c; \theta)
    \label{eq:cfg_ours}
\end{equation}

\vspace{-0.1cm}
\section{Experimental Setup}
\label{sec:exp_setup}
\vspace{-0.05cm}
\subsection{Data Preparation}
The training speech corpus used in our experiments is LibriTTS~\cite{zen2019libritts}. To simulate various acoustic conditions, we augment this corpus with additional noise from the  WHAM!~\cite{wichern2019WHAM} dataset using a randomly selected SNR ranging from -5 to 10dB.  We use RIR-NOISES ~\cite{ko2017study} to simulate reverberation effects. We randomly select utterances from a different speaker within the LibriTTS dataset as the interfering speech, and we mix the main and interfering speech with SNR values ranging from 1 to 10dB to ensure that the main speech has higher energy. We prepare two test sets.  The in-domain test set consists of the LibriTTS test set, augmented with WHAM! and RIR-NOISES test data with the same SNR range as the training set. To assess out-of-domain performance, we simulate the VCTK~\cite{Veaux2017CSTRVC} speech dataset and the TUT~\cite{mesaros2016tut} noise evaluation dataset with SNR ranging from -5 to 10dB. During experiments, we resample all waveforms to 16000Hz. 

\vspace{-0.05cm}
\subsection{System Configuration}
The probabilities $P^{N}$, $P^{R}$, $P^{IS}$ in CMSP strategy are set to $20\%$ and $P^C=40\%$. The binary control signal $c$ is represented by a one-dimensional tensor containing either all zeros or all ones, determining which task to be performed. The acoustic model is composed of a transformer encoder with 4 layers, 16 heads, and a dimensionality of 1024 as the backbone, along with two small transformer encoders as the speaker encoders, each with 2 layers, 2 heads, and a dimensionality of 80. The duration predictor, which is the same across all baselines and our models, is a transformer encoder with 4 layers, 16 heads, and a dimensionality of 1024. We follow~\cite{zhang2024covomix} to set all hyper-parameters. We train all models with the Adam optimizer for 100 epochs, and the batch sizes for the acoustic model and duration predictor are 8 and 24 respectively. We train a HiFiGAN~\cite{kong2020hifi} vocoder on LibriTTS for 300k steps to reconstruct waveforms from mel-spectrograms. All experiments are conducted on four NVIDIA A10 GPUs.

\subsection{Baselines}
We benchmark our method towards three baselines: the VoiceBox model, the VoiceBox model with a universal SE model to process the speech prompts
~\cite{zhang2023toward}, and the VoiceBox Model with the MSD strategy.
All baselines are reimplemented using public code\footnote{https://github.com/lucidrains/VoiceBox-pytorch} and are trained with the same backbone and data as our system.  

\subsection{Evaluation Metrics}
The objective metrics are Speaker Similarity (SIM) and Mel Cepstral Distortion (MCD). We extract speaker embeddings using WavLM~\cite{chen2022large,chen2022wavlm} and calculate the cosine similarity between embeddings of the generated speech and the clean prompt reconstructed by Vocoder. For the background removal task, MCD is calculated against the clean ground truth, while for the background preservation task, it is calculated against the noisy ground truth, which retains the same acoustic information as the prompt. 

For the subjective evaluation, we conduct two Mean Opinion Score (MOS) listening tests. 15 listeners are asked to rate each utterance on a scale from 1 to 5. The conventional MOS~\cite{international1996methods} is used to assess speech quality for the background removal task, while the Background Similarity (B-MOS) score is used for the background preservation task. Each listener evaluates 10 utterances for each environmental scenario.

\section{Results and Analysis}
\label{sec:result}

\subsection{Evaluation on Background Removal}
Table \ref{tab:removal} compares the performance of background removal of our method and other baselines in various scenarios. We first observe that the VoiceBox baseline shows unstable results with different prompts. It can clone the target speaker's voice in scenarios with reverberation and interfering speakers, but it is less robust for environmental noise. The addition of the SE model or MSD strategy results in a cleaner background of the generated speech, but with a loss of timbre similarity, especially in the clean scenario.

\begin{figure}
    \centering
\includegraphics[width=0.86\linewidth]{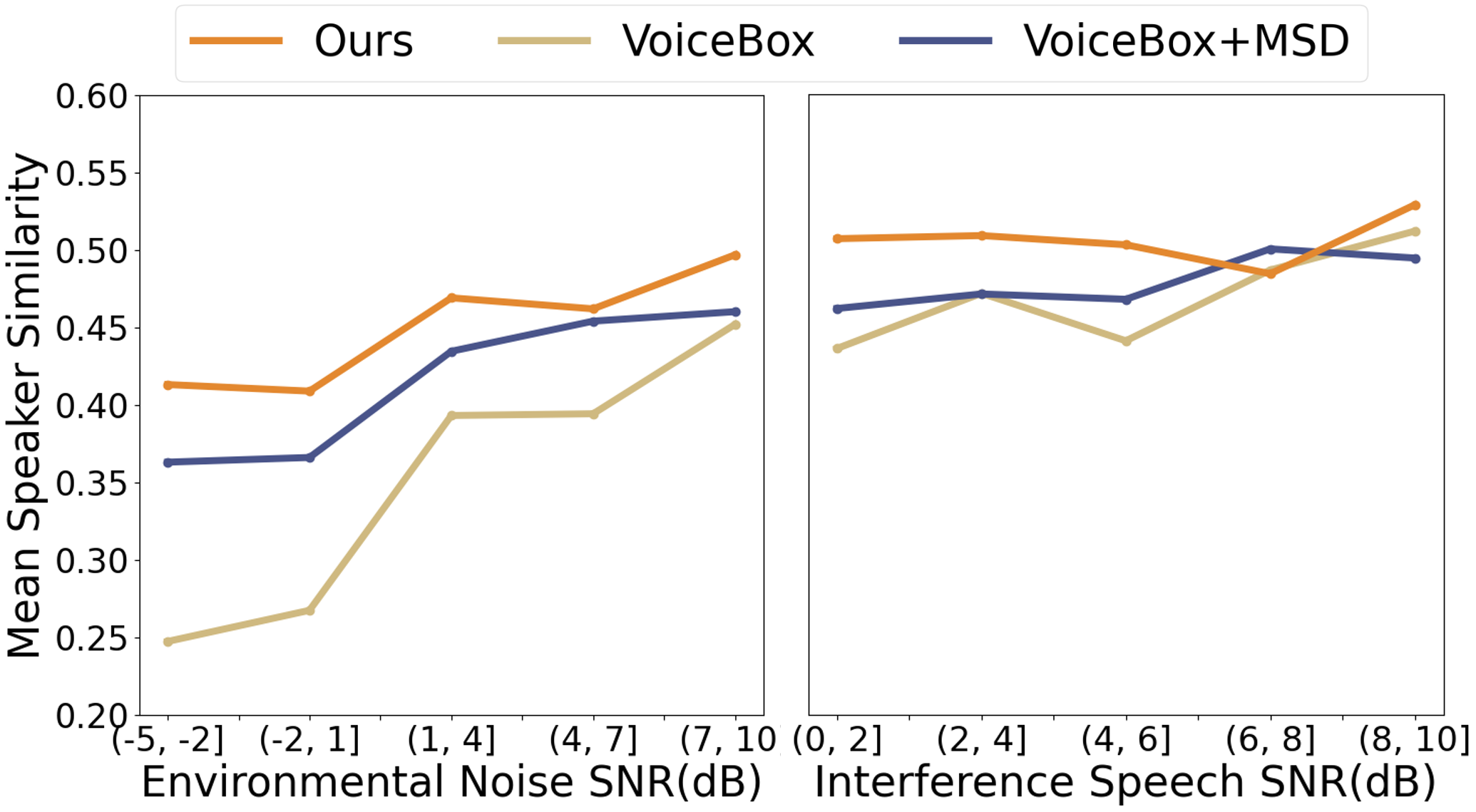}
    \caption{Mean Speaker Similarity Comparison by SNRs for Each Model}
    \label{fig:snr}
    \vspace{-0.4cm}
\end{figure}

Our method shows comparable performance compared with the VoiceBox model in the clean condition, and achieves obviously higher speaker similarity and speech quality in all noisy, reverberant, and interfering speaker scenarios compared to the baseline augmented with an extra SE model and the baseline using MSD strategy. Moreover, Figure \ref{fig:snr} also demonstrates the robustness of our model across different SNR levels. It is observed that the intensity of environmental noise has a noticeable impact on speaker similarity for different SNR levels, whereas the intensity of interfering speech has minimal effect. Our approach always shows the best robustness across all SNR levels.

\subsection{Evaluation on Background Preservation}
\begin{figure}
    \centering
\includegraphics[width=0.86\linewidth]{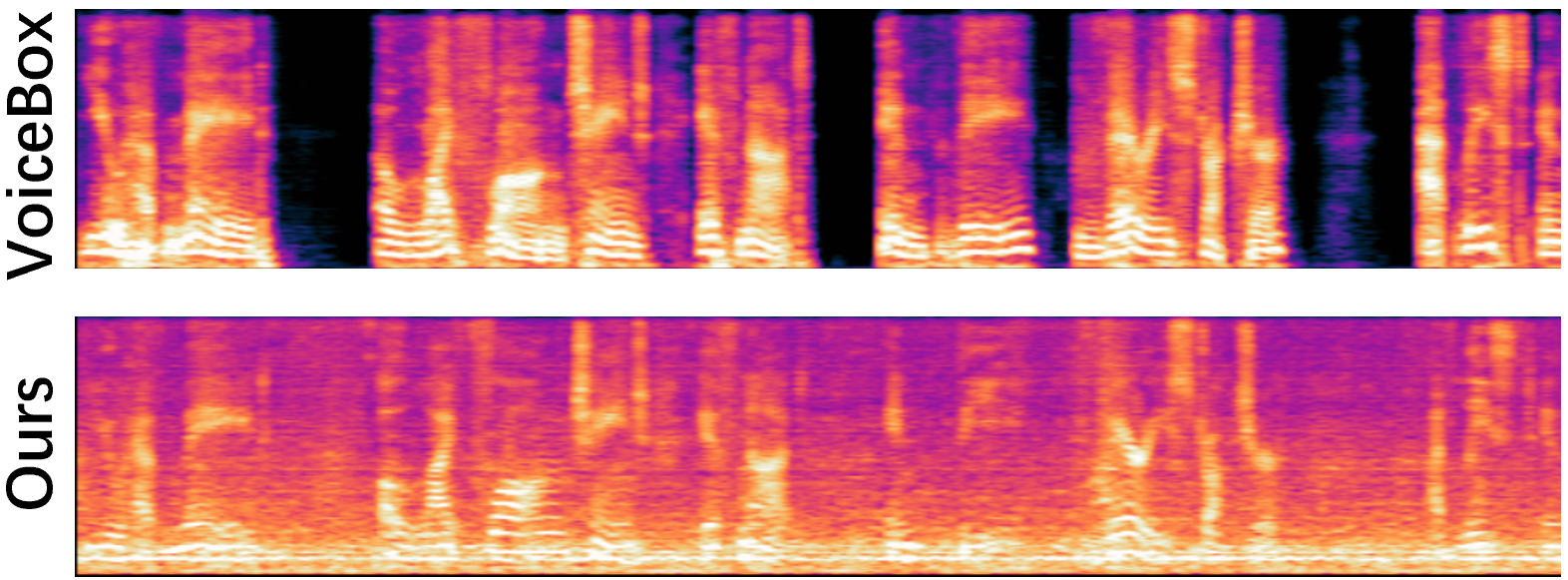}
    \caption{Mel-spectrogram Comparison for Generated Speech with Preserved Background Noise among Models}
    \label{fig:background-noise}
        \vspace{-0.4cm}
\end{figure}
Table \ref{tab:preserve} compares the performance of background preservation of our method and the VoiceBox baseline. The VoiceBox baseline can preserve most of the reverberation and some degree of background noise,  but it can not generate interfering speeches. However, the VoiceBox baseline does not distinguish between prompts with and without background in its modeling process. The generated mel-spectrograms are illustrated in Figure \ref{fig:background-noise}, and it is observed that the generated background is only present during active speech segments, and the background will pause when the speaker pauses. In contrast, our proposed method, modeling prompts with and without background differently, is able to generate a more consistent and continuous background, indicating the strong background preservation capabilities of our proposed model.

However, with interfering speakers, our model is slightly worse than VoiceBox only on the MCD metric. This discrepancy arises because no transcription is provided to guide the generation of the interfering speech, resulting in semantic misalignment between the generated and ground truth backgrounds, thereby leading to a higher MCD value. In order to validate the timbre similarity of the generated interfering speakers, we use a pre-trained speech separation (SS) model~\cite{luo2023music} to process the generated speech. 
Despite the presence of some inaccuracies in the SS model's results, the SIM between the interference speaker in the prompt and in the generated speech is 0.20, indicating that our model successfully clones the interfering speaker's characteristic and there is still room for improvement.

\subsection{Out-of-Domain Generalibility}

We evaluated model performance on an out-of-domain dataset with environmental noise. As shown in Table \ref{tab:removal}, our method outperforms VoiceBox and VoiceBox+MSD in the background removal task. However, the speaker similarity is lower compared to the VoiceBox+SE, because the latest SE model~\cite{zhang2023toward} exhibits very strong robustness to unseen noise with much more noise types data. 
When coupled with this SE module to first enhance the speech prompt, our method can also further improve speaker similarity in handling unseen noises. 

Moreover, Table \ref{tab:preserve} illustrates that our model maintains a higher level of background preservation than baseline system for unseen noises in out-of-domain scenarios, which further demonstrates the robustness and generalizability of our proposed method.

\subsection{Ablation Study}
To further understand the impact of training strategy and speaker encoder, we conducted an ablation study focusing on speaker similarity in the background removal task, with results shown in Table \ref{tab:abalation-removal}. The CMSP strategy outperforms the MSD strategy in terms of speaker similarity in all scenarios, and the dual speaker encoders perform better than the single encoder of the same architecture. The application of both types of speaker encoders enhances speaker similarity, and the dual encoder shows more significant improvement.

Furthermore, Table \ref{tab:abalation-preservation} compares the performance of systems with different speaker encoder configurations for the background preservation task.  The results show that the dual speaker encoder benefits background information preservation, especially in scenarios involving interference speakers. This further demonstrates the effectiveness of modeling different tasks separately in the model training process. Although a single-speaker encoder can still function, it tends to produce less natural speech when dealing with interfering speeches. 

\begin{table}[t]
    \centering
    \caption{Ablation Study of Training Strategy and Speaker Encoder Architecture on Speaker Similarity for Background Removal}
    \label{tab:abalation-removal}
    \begin{tabular}{c|cc|cccc}
   \toprule
        ID & SpkEncoder & Strategy & Clean & Noise & Reverb & Interference  \\ \midrule
        R1 & No & MSD & 0.54 & 0.39 & 0.47 & 0.46  \\
        R2 & No & CMSP & 0.57 & 0.41 & 0.49 & 0.50  \\
        R3 & Single & MSD & 0.56 & 0.41 & 0.49 & 0.47  \\
        R4 & Single & CMSP & 0.58 & 0.42 & 0.50 & 0.50  \\
        R5 & Dual & CMSP & \textbf{0.59} & \textbf{0.44} & \textbf{0.51} & \textbf{0.50}  \\ \bottomrule
    \end{tabular}
        \vspace{-0.3cm}
\end{table}

\begin{table}[t]
    \centering    
    \caption{Ablation Study of Speaker Encoder Architecture on Background Similarity for Background Preservation}
    \label{tab:abalation-preservation}
    \begin{tabular}{c|cc|ccc}
   \toprule
        ID & SpkEncoder & Strategy & Noise & Reverb & Interference  \\ \midrule
        P1 & No & CMSP & 3.10 & 2.98 & 2.35  \\
        P2 & Single & CMSP & 3.67 & 3.35 & 2.48  \\
        P3 & Dual & CMSP & \textbf{3.78} & \textbf{3.51} & \textbf{3.09}  \\
        \bottomrule
    \end{tabular}
            \vspace{-0.3cm}
\end{table}

\section{Conclusion}
In this work, we propose a novel zero-shot TTS model with the Controllable Masked Speech Prediction strategy and the Dual Speaker Encoder, which can effectively achieve the background removal and background preservation tasks in one integrated model. Experiments demonstrate the model's capability to handle prompts in various acoustic conditions in both in-domain and out-of-domain scenarios. In the future, we will continue improving the quality of generated backgrounds and further expand into more scenarios such as music. 

\section*{Acknowledgment}
This work was supported in part by China NSFC projects under Grants 62122050 and 62071288, in part by Shanghai Municipal Science and Technology Commission Project under Grant 2021SHZDZX0102.
\bibliographystyle{IEEEtran}
\bibliography{IEEEexample}

\end{document}